\documentclass[aps,prl,preprint,groupedaddress]{revtex4-1}
\usepackage{graphicx}
\usepackage{graphics}
\usepackage{epsfig}
\usepackage{epstopdf} 

\begin{document}

\title{An operational definition of the 100 second blocking temperature T$_{\mathrm{B}100}$ for single molecule magnets}

\author{Rasmus Westerstr\"om}
\affiliation{Physik-Institut, Universit\"at Z\"urich, Winterthurerstrasse 190, CH-8057 Z\"urich, Switzerland}
\affiliation{Department of Physics and Astronomy, Uppsala University, Box 516, S-751 20 Uppsala, Sweden}

\author{Alexey Popov}
\affiliation{Leibniz Institute of Solid State and Materials Research, Dresden, D-01069 Dresden, Germany}

\author{Thomas Greber}
\affiliation{Physik-Institut, Universit\"at Z\"urich, Winterthurerstrasse 190, CH-8057 Z\"urich, Switzerland}

\begin{abstract}

An important figure of merit for the performance of single-molecule magnets (SMMs) is the 100 s blocking temperature T$_{\mathrm{B}100}$. It is the temperature at which the remanence or zero field relaxation time is 100 seconds. If there is more than one relaxation process of the magnetisation, the determination of the relaxation times may, however, become ambiguous. Here we propose an operational definition for the zero-field magnetic relaxation times from which T$_{\mathrm{B}100}$ may be determined. This definition allows for a direct comparison of the performance of different samples independent of the details of the relaxation processes involved in the demagnetization.

\end{abstract}

\maketitle


Single-molecule magnets (SMMs) \cite{sessoliNat93} represent a class of compounds that exhibits an intrinsic magnetic bistability at low-temperatures. These systems can exhibit magnetic hysteresis and remanence similar to macroscopic magnets, as well as quantum phenomena \cite{Gatteschi03}. They show potential for application in quantum-computing \cite{leuenbergerNat01}, spintronics \cite{bogani08mNat}, and might represent the ultimate limit for data-storage. A limitation of these systems is that remanence is only observed for a certain time and below a certain temperature. Synthesizing SMMs that exhibit magnetic bistability at higher temperatures is, therefore, desirable and quantitative measures for comparison and benchmarking of different samples are imperative. Unlike large magnetic systems, where a characteristic temperature like the Curie or the N\'eel temperature characterize a phase transition, in superparamagnetic systems a blocking temperature has to be defined. This is more difficult since the blocking temperature does not represent a thermodynamic equilibrium, but is a kinematic quantity.  
The so-called magnetic blocking temperature T$_{\mathrm{B}}$, corresponding to the temperature where the decay time of the magnetization is in the order of the measurement time, is frequently used as a figure of merit for the performance of an SMM. A quantitative definition has been proposed by Gatteschi \textit{et al}, where the blocking temperature T$_{\mathrm{B}100}$ corresponds to the temperature  at which the system exhibits 100 s relaxation time \cite{GatteschiBook06}. Remanence decay times on this timescale are usually determined by fully magnetizing the sample, ramping the field to zero and then recording the magnetization as a function of time. In the simplest case, a single exponential function 
\begin{equation}
M(t) \propto \exp(-t/\tau)
\label{eq1}
\end{equation}
is fitted to the data and the relaxation time $\tau$ is extracted. If there is one relaxation process only, if $\tau$ depends on the temperature only, and if relaxation would start after reaching zero field only, this would be a straight forward way to find a blocking temperature for a given decay time, of e.g. 100 s, where after 100 s about 63\% of the initial magnetisation is lost. The relaxation rate $1/\tau$ does, however, depend on the field, and the magnetisation decay during the ramping of the magnetic field. Depending on the complexity of the relaxation process, decay-avalanches or isotope dependencies may be important and the kinetics of the decay may be more involved. For many samples it is empirically found that a single exponential as described in Equation \ref{eq1} may not describe the observations. For the N$_2^{3-}$ radical-bridged Tb and Dy complexes \cite{RinehartJACS2011}, the Dy$_2$ScN@C$_{80}$ \cite{westerstromPrb14} and the DySc$_2$N@C$_{80}$ \cite{westerstromJACS} endofullerene a double exponential was pragmatically used to extract decay times, where however, the faster \cite{RinehartJACS2011} and the slower \cite{westerstromPrb14} decay times have been used for the determination of the 100 s blocking temperature. 
Furthermore, the double exponential method as applied so far did not account for the initial ratio of the two decay channels $\frac{M_A}{M_B}$ (see Equation \ref{eq2}), nor the decay during ramp-down of the magnet from saturation to zero field. These two examples interpolate a T$_{\mathrm{B}100}$ temperature from observed decay rates.
There are also schemes found in the literature, where extrapolation is used for the demagnetisation rates, a procedure that may predict misleading blocking temperatures.
In a recent study of Er$_2$-COT compounds \cite{leRoyJACS}, the T$_{\mathrm{B}100}$ was determined by extrapolating the ac magnetic life-times to lower temperatures. If we apply this method to the data of Dy$_2$ScN@C$_{80}$ \cite{westerstromPrb14}, we obtain a T$_{\mathrm{B}100}$ of 5.5 K. 
We consider this method to be inappropriate since the onset of importance of other decay mechanisms at lower temperatures may make the extrapolation obsolete.
Apparently, different blocking temperatures can be extracted for the same sample, which makes a direct comparison of the performance in terms of blocking temperatures for different samples impossible. This ambiguity calls for a convention on how to determine T$_{\mathrm{B}100}$.

\begin{figure}[t]
\begin{center}
\includegraphics[width=12cm]{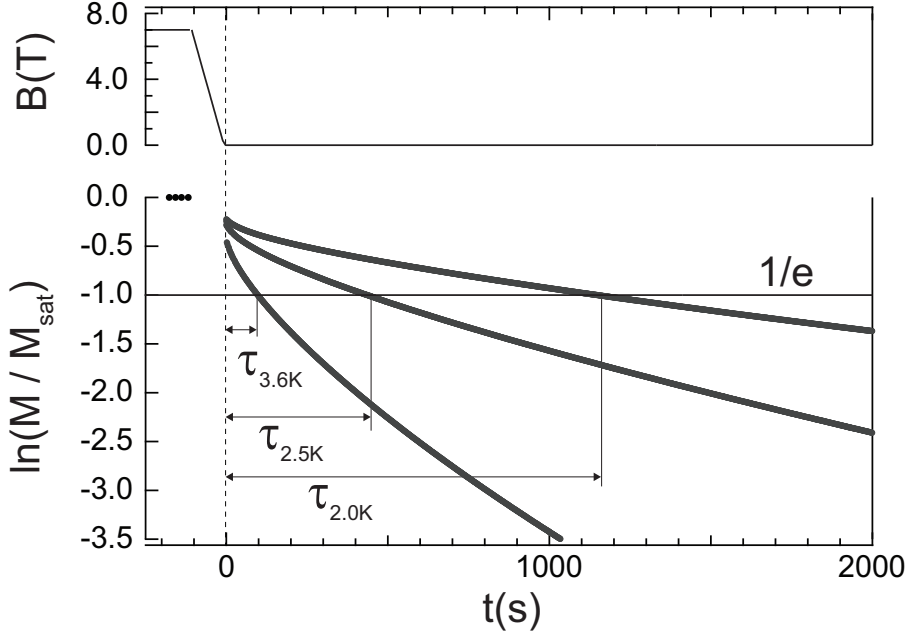}
\caption{Relaxation curves recorded from Dy$_2$ScN@C$_{80}$ for different temperatures after saturation at 7 T. The magnetization is normalized to the saturated value at 7 T. The relaxation time is defined as the time at which the magnetization has decayed to 1/$e$ of the saturated magnetization.}
\label{fig1}.
\end{center}
\end{figure}

Here we propose a simple method of extracting the remanence (zero field) demagnetization times for the determination of the 100 s blocking temperature. This method uses no extrapolation and is  independent of the details of the relaxation processes: We measure the decay time as the time that elapses until the saturation magnetisation drops below $\frac{M_{sat}}{e}$, where $M_{sat}$ is the saturation magnetisation and $e$ the Euler number. The procedure is illustrated in Figure \ref{fig1} for the case of the endohedral SMM Dy$_2$ScN@C$_{80}$ \cite{westerstromPrb14}. First the system is cooled down from room-temperature to the investigated temperature in zero magnetic field. The sample is then fully magnetized at 7 T, after which the external field is ramped down in 100 seconds, i.e. at a rate of 70 mTs$^{-1}$. The time at which the field has reached zero defines the starting-point, t=0, for the recording of the relaxation curves. The relaxation time $\tau$ is then defined as the time when the magnetization has decayed to 1/$e$ of its saturation value $M_{sat}$ measured at 7 T. 
Note that the normalization of M(t) with M$_{sat}$ and not with M(t=0) further decreases the relaxation time for a given temperature, however for any technological application one is interested in the decay time of the initial magnetisation.
By measuring relaxation times for different temperatures, we then determine T$_{\mathrm{B}100}$ by interpolation. 

\begin{figure}[t]
\begin{center}
\includegraphics[width=12cm]{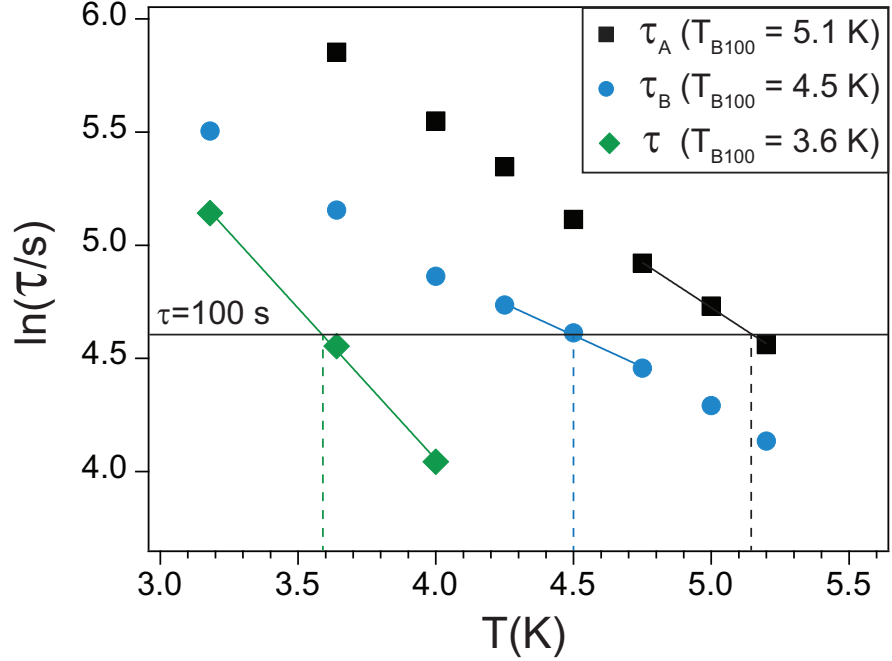}
\caption{Zero field relaxation times of the Dy$_2$ScN@C$_{80}$ sample from Ref.\cite{westerstromPrb14} extracted by fitting a double exponential to the relaxation curves($\tau_{\mathrm{A}}$ and $\tau_{\mathrm{B}}$), and applying the new definition in Fig. \ref{fig1} for the $\frac{M_{sat}}{e}$ times ($\tau$). The 100 s blocking temperatures are determined from the intersection of the different decay time curves and the line corresponding to 100 s. The value of 3.6 K for T$_{\mathrm{B}100}$ corresponds to the new definition.}
\label{fig2}.
\end{center}
\end{figure}

Figure \ref{fig2} displays the $\frac{M_{sat}}{e}$ relaxation times and the decay times $\tau_A$ and $\tau_B$ extracted by fitting a double exponential to the relaxation curves:
\begin{equation}
M(t) = M_A \exp(-t/\tau_A)+M_B \exp(-t/\tau_B),~~t>0
\label{eq2}
\end{equation}

Depending on the method used to extract the relaxation times, T$_{\mathrm{B}100}$ varies between 5.1~K and 3.6~K for the same sample, where the value of 3.6~K is the result of the new procedure for the blocking temperature determination. The data in Figure 2  demonstrates the need for a precisely defined method to determine the blocking temperature T$_{\mathrm{B}100}$. It has to be mentioned that the newly proposed method depends on the magnetic field sweep rate which therefore has to be defined as well if different samples shall be compared. It appears to be practical to use a ramp down time of 100 s as it is used here for the determination of T$_{\mathrm{B}100}$. 
Last but not least we would to emphasize that the proposed method does not require time consuming measurements of the slow relaxation rate.

In summary we propose an operational definition of the figure of merit the 100 seconds blocking temperature T$_{\mathrm{B}100}$ of superparamagnetic single molecule magnets.
  
\end{document}